# Future optical detectors based on Al superconducting tunnel junctions

G. Brammertz[*a], P. Verhoeve[a], D. Martin[a], A. Peacock[a], R. Venn[b]

[a]Science Payload and Advanced Concepts Office, ESA/ESTEC, P.O. Box 299, 2200AG Noordwijk, The Netherlands.

[b]Cambridge MicroFab Ltd., Broadway, Bourn, Cambridgeshire CB3 7TA, UK.

## ABSTRACT

Superconducting tunnel junctions are being developed for application as photon detectors in astronomy. We present the latest results on the development of very high quality, very low critical temperature junctions, fabricated out of pure Al electrodes. The detectors are operated at 50 mK in an adiabatic demagnetisation refrigerator. The contacts to the top and base electrodes of these junctions are fabricated either out of Nb or Ta, which has strong implications on the loss time of the quasiparticles. The Nb contacted junctions show quasiparticle loss times varying between 5 and 80 μsec, depending on the device size. The bias range of the Nb-contacted junctions is limited to the range 0-100 μV, because of the set-in of strong non-equilibrium quasiparticle multiplication currents at higher bias voltages. The Ta-contacted junctions, on the other hand, show quasiparticle loss times in excess of 200 μsec. These long loss times lead to very strong quasiparticle multiplication, which prevents the stable biasing of the junctions even at very low bias voltages. Junction fabrication and characterisation are described, as well as the response of the detectors to monochromatic light with wavelengths varying from 250 to 1000 nm. The energy resolution of the detectors is discussed.

## 1. INTRODUCTION

For a number of years superconducting tunnel junctions (STJs) have been developed as photon counting spectrometers for application in optical astronomy [1]. STJs consist of two superconducting electrodes separated by a thin insulating tunnel barrier (SIS-junction). Figure 1 shows an example of such a junction based on Ta-Al electrodes separated by a thin Al oxide barrier. A magnetic field applied in parallel to the barrier suppresses all zero voltage Josephson currents in the junction. As a photon is absorbed in one of the superconducting electrodes a certain number of Cooper pairs are broken into high-energy quasiparticles. After a series of quasiparticle energy down-conversions and subsequent Cooper pair breakings by the released phonons, the energy of the initial photon is transformed into a cloud of low energy quasiparticles and phonons. At the end of the process an initial number of quasiparticles $Q_0=E/1.7\Delta_g$ is created in the superconductor [2], where E is the energy of the absorbed photon and $\Delta_g$ is the energy gap of the absorbing superconducting electrode. As the energy gap of a typical superconductor is of the order of a meV, every eV of photon energy will create several hundreds or thousands of quasiparticles. Upon application of a small bias voltage across the junction, these quasiparticles will start tunneling across the thin insulating barrier thereby creating an electrical current. Once the quasiparticles have tunneled, they have the possibility to back-tunnel into the initial electrode, a process, which also creates a current in the same direction as the tunnel process [3]. There is thus effectively an amplification of the output signal by the number of tunnel events a single quasiparticle undergoes. The current created by the tunnel processes can be integrated and the collected charge is correlated to the energy of the initial photon absorbed in the electrode. In this way, every single photon absorbed in the electrode can be detected and analyzed.

Next to photon counting capabilities, time resolution of the order of a microsecond and spatial resolution (for STJs arranged in an array form), the STJs possess an intrinsic full width at half maximum (FWHM) energy resolution given by:

$$\Delta E = 2.355\sqrt{1.7\Delta_g (F+G)E} \ . \qquad (1)$$

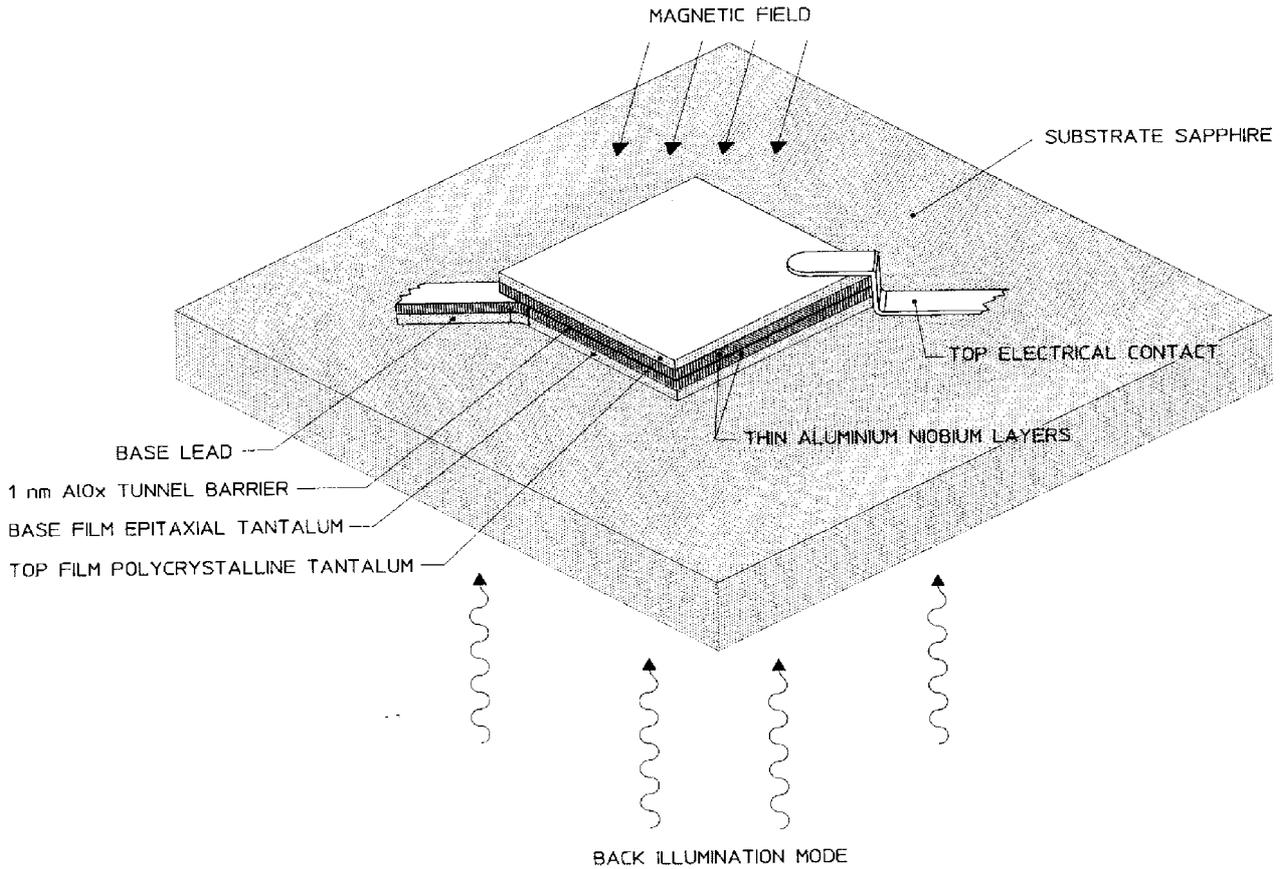

**Figure 1.** Schematic of a Ta-Al superconducting tunnel junction used as photon detector.

Here, F is the Fano factor and G is associated with the variance in the tunnel term. The Fano factor F represents the resolution broadening due to statistical variations in the initial number of created quasiparticles by the photon absorption process. Its value was determined to be equal to 0.2 for simple low critical temperature superconductors, like the ones used for the fabrication of STJs [4,5]. The tunneling term G represents statistical variations in the number of tunnel events a quasiparticle undergoes before it is finally lost. G is equal to 1+1/<n>, where <n> is the mean number of tunnel events per quasiparticle [6,7]. Table 1 summarizes the characteristics of some basic superconductors as well as the expected best achievable resolution according to equ. (1) at 2.48 eV, 1 keV and 6 keV.

| Material | Energy gap $\Delta_g$ | Critical temperature $T_C$ | $\Delta E$ at 2.48 eV | $\Delta E$ at 1 keV | $\Delta E$ at 6 keV |
|---|---|---|---|---|---|
| | (µeV) | (K) | (eV) | (eV) | (eV) |
| Niobium (Nb) | 1550 | 9.3 | 0.208 | 4.2 | 10.2 |
| Vanadium (V) | 820 | 5.4 | 0.15 | 3.0 | 7.5 |
| Tantalum (Ta) | 700 | 4.5 | 0.14 | 2.8 | 7 |
| Aluminium (Al) | 180 | 1.2 | 0.07 | 1.4 | 3.5 |
| Molybdenum (Mo) | 139 | 0.915 | 0.06 | 1.25 | 3.1 |
| Hafnium (Hf) | 19.4 | 0.128 | 0.023 | 0.47 | 1.15 |

**Table 1.** Energy gap, critical temperature and best expected FWHM energy resolutions at three different photon energies for different superconducting materials.

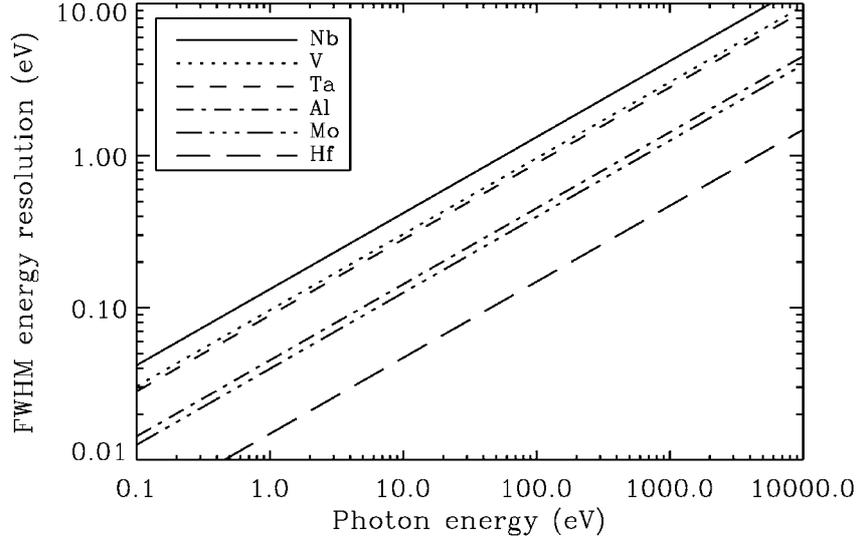

**Figure 2.** Expected ideal energy resolution as a function of incoming photon energy for six different superconducting materials.

The expected best energy resolution is shown graphically in figure 2 for the six different superconductors of table 1 as a function of the incoming photon energy. The ideal resolution achievable varies as the square root of the energy gap of the superconductor. In the optical, this optimum energy resolution performance has already been demonstrated with Ta-based junctions, where intrinsic resolutions of 0.15 eV at 2.48 eV ($\lambda$ = 500 nm) are routinely achieved [1]. 10 x 12 pixel arrays of this kind of Ta-based junctions are currently being operated at the William Herschel Telescope in La Palma [8]. For a future generation of detectors, the energy resolution should be improved to values below 0.1 eV at 2.48 eV. Therefore, our approach is to fabricate the junctions out of the lower $T_C$ materials, like Al and Mo. In the following, we present the characteristics of our pure Al based ($T_C$ = 1.2K), high quality STJs and present optical photon detection experiments made with these junctions operated at a temperature of 50 mK.

## 2. DETECTOR FABRICATION

The pure Al junctions are fabricated by Cambridge MicroFab Ltd. In a UHV system, a polycrystalline Al film is DC-sputtered at liquid nitrogen temperature onto a Kyocera r-plane sapphire substrate. This base Al film has a thickness of approximately 100nm and a residual resistance ratio of about 10. Without removing the film from the UHV system, it is then oxidized in a 99.999% pure oxygen environment in order to grow the approximately 1nm thick aluminum oxide layer. On top of this insulating layer is then deposited the top electrode of the junction, a 50 nm thick polycrystalline Al film. The completed tri-layer is then removed from the UHV system and a 1 to 4 nm thick natural oxide layer forms on top of the tri-layer. UV lithography is used to pattern the resists for the subsequent steps. The etch through the complete tri-layer is done with an MIT acid etch. This step is followed by the mesa etch, which only etches through the top film and the barrier, but leaves the base film as far as possible untouched. This mesa etch is done using a neutralized ion beam miller. The complete wafer is then covered by a 350 nm thick, insulating silicon oxide (SiOx) film. The SiOx film is reactively sputtered from an intrinsic Si target into an Ar environment with 10 % of O. Vias are formed through the insulating SiOx film where appropriate using plasma etching with an $SF_6$ gas. As a final step the 400 nm thick Nb or Ta top contacts and/or base plugs are sputtered and patterned using a two-layer resist lift-off procedure. This Nb or Ta plugs in the base film contact prevent the out-diffusion of quasiparticles out of the junction area, which strongly reduces quasiparticle losses.

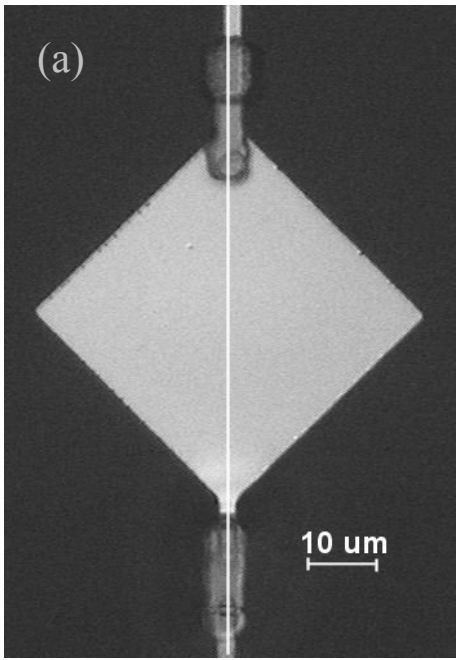 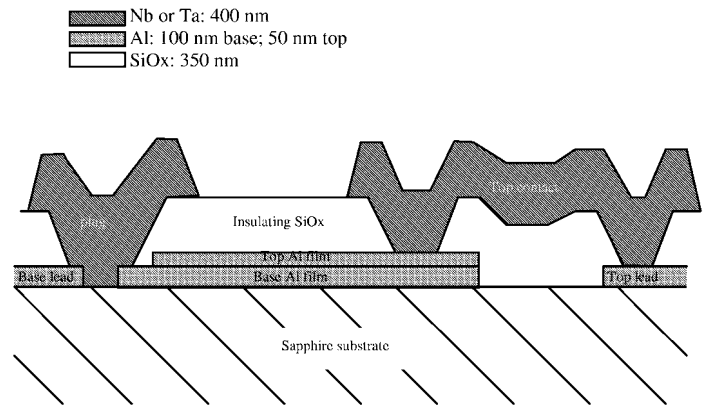

**Figure 3. (a)** Nomarski microscope image of a 40 μm side length Al STJ with a Nb plug in the base contact and a Nb top contact. The vertical white line indicates the position of the cross section through the device, which is schematically shown in **(b).**

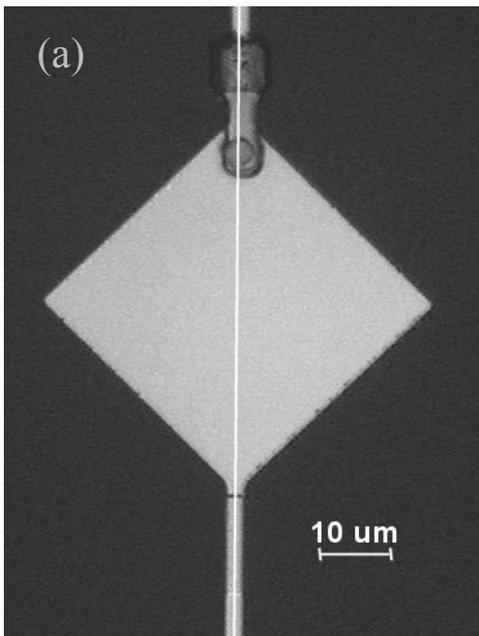 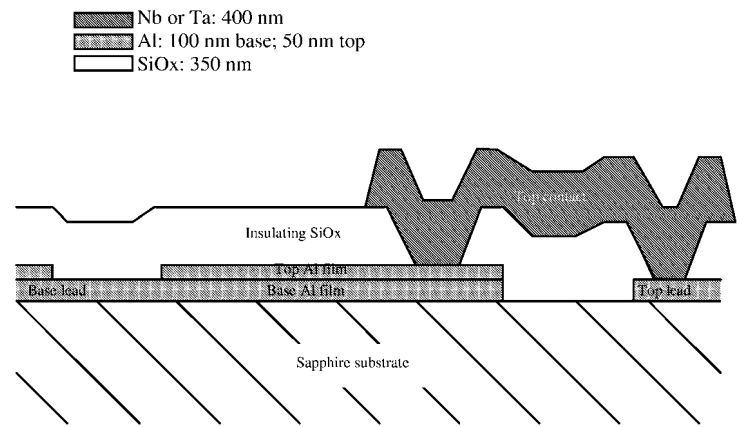

**Figure 4. (a)** Nomarski microscope image of a 40 μm side length Al STJ without a Nb plug in the base contact and a Nb top contact. The vertical white line indicates the position of the cross section through the device, which is schematically shown in **(b)**.

Two different chip layouts are available: one chip with junctions having a plug in the base electrode contact and with device sizes varying from 10 to 100 μm and one chip with all 40 μm side length junctions, of which half of the junctions have a plug in the base electrode contact lead and half of the junctions do not have such a plug.

Figure 3(a) shows a Nomarski microscope image of a fully processed, 40 μm side length Al STJ with a Nb plug in the base contact and a Nb top contact. The vertical white line indicates the position of the cross section through the device, which is schematically shown in Figure 3(b). Figure 4 shows the corresponding images for a junction that does not posses such a plug in the base electrode contact.

## 3. EXPERIMENTAL SET-UP

The junctions are cooled in a two-stage adiabatic demagnetization refrigerator (ADR) [9] possessing a single magnet and two concentric salt pill reservoirs. The high temperature paramagnetic material is Gadolinium Gallium Garnet (GGG) whereas the low temperature paramagnetic salt is Ferric Ammonium Alum (FAA). Both pills are suspended from the 4K He reservoir via Kevlar strings. The system has a base temperature of 50 mK and a hold time below 100 mK in excess of 10 hours. The sample space is magnetically shielded from the pill system in order to avoid magnetic flux trapping in the junctions. Inside the magnetically shielded sample space a superconducting magnet is able to apply a magnetic field in the plane parallel to the junctions barrier in order to suppress the zero voltage Josephson current.

The junctions are coupled via an optical fiber to a Xenon lamp in combination with a double grating monochromator, able to produce photons having a wavelength varying from 250 to 1000 nm. The illumination of the junctions is made through the back of the chip via the sapphire substrate.

For optical photon detection experiments the junctions are read out through a room-temperature charge-sensitive preamplifier followed by a shaping stage. The charge sensitive preamplifier integrates the current pulse created by the photo-absorption process in the junction. The output signal of the preamplifier is then fed into the shaping stage, consisting of two bipolar semi-gaussian shaping filters with fixed center frequencies of respectively 5 and 33 kHz. In parallel to the photo-signal, an electronic pulser can be fed into the preamplifier, in order to measure the electronic noise.

## 4. NB-CONTACTED AL JUNCTIONS

We will first present the properties of our Al junctions with Nb top contacts and plugs. First the IV-characteristics are discussed, which give a good indication of the quality of the junctions and then the response of the junctions to optical photons is presented.

### 4.1. I-V characterisation

Figure 5(a) shows the complete IV-curve of a 20 μm side length junction taken at 50 mK. A magnetic field of approximately 30 Gauss was applied parallel to the junction barrier in order to suppress the zero voltage Josephson current to an acceptable level. The top and base leads of this junction are 6 μm wide in order to prevent breakdown in the leads at high current densities. In the figure one can identify the sum-gap of the electrodes equal to 350 μeV. The almost vertical dashed line is a fit to the sum-gap region of the curve. Assuming a symmetrical lay-up, one deduces an energy gap equal to 175 μeV in both electrodes, which is reasonably close to the bulk energy gap of Al equal to 180 μeV. The normal state resistance $R_n$ of the junction can also be determined from the figure, and is equal to 1.75 Ω for this 20 μm side length junction. This gives a normal state resistivity $\rho_{nn}$ equal to 7 μΩ cm$^2$.

Figure 5(b) shows the sub-gap regimes of three different junctions with side lengths of respectively 30, 50 and 70 μm. The IV-curves were acquired at a temperature of 50 mK. A magnetic field of approximately 30 Gauss was applied in parallel to the junction in order to suppress the Josephson current below 10 nA. The dashed lines are fits to the dynamical resistances $R_d$ in the sub-gap domain, which is the region in which the junctions are biased for photon detection experiments. The sub-gap currents in the operational bias voltage regime (~30-100 μV) are proportional to the area of the junction and are equal to 260 fA per μm$^2$ of junction area at a bias voltage of 50 μV. This shows that the residual currents in the bias domain are leakage currents arising from very small pinholes in the insulating barrier distributed homogeneously over the area of the junction. The dynamical resistivity $\rho_d$ of the Al based junctions can be derived and is equal to 1.9 Ω cm$^2$. This yields a quality factor Q = $\rho_d/\rho_{nn}$ = 2.7 10$^5$, which underlines the very good quality of our Al junctions.

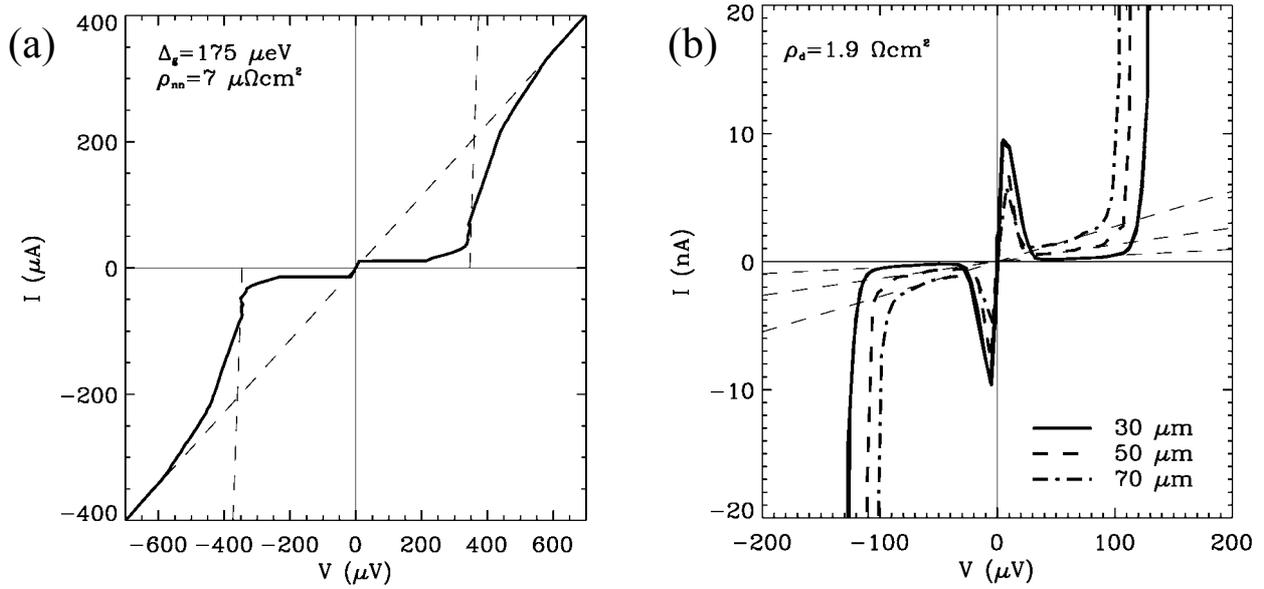

**Figure 5.** IV-curves of Nb-plugged Al-based junctions taken at a temperature of 50 mK. The applied parallel magnetic field is of the order of 30 Gauss. **(a)** Full IV-curve of a 20 μm side-length STJ. The dashed lines are fits to the sum-gap of the two electrodes and the normal resistance of the junction. **(b)** IV-curves of the sub-gap region of three Nb-plugged junctions with side-lengths of 30, 50 and 70 μm. The thin dashed lines are fits to the dynamical resistances in the sub-gap regime.

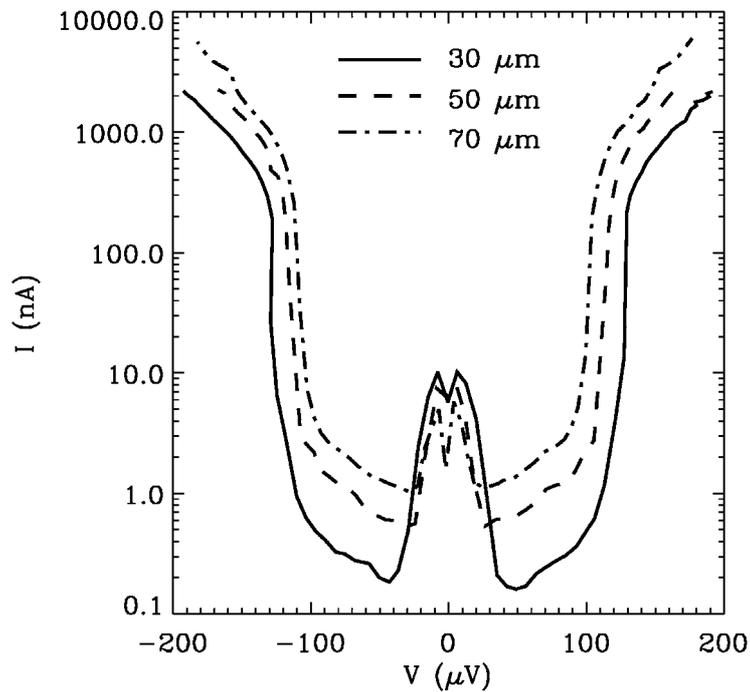

**Figure 6.** IV-curves of the 30, 50 and 70 μm side-length junctions with Nb plugs on a logarithmic scale.

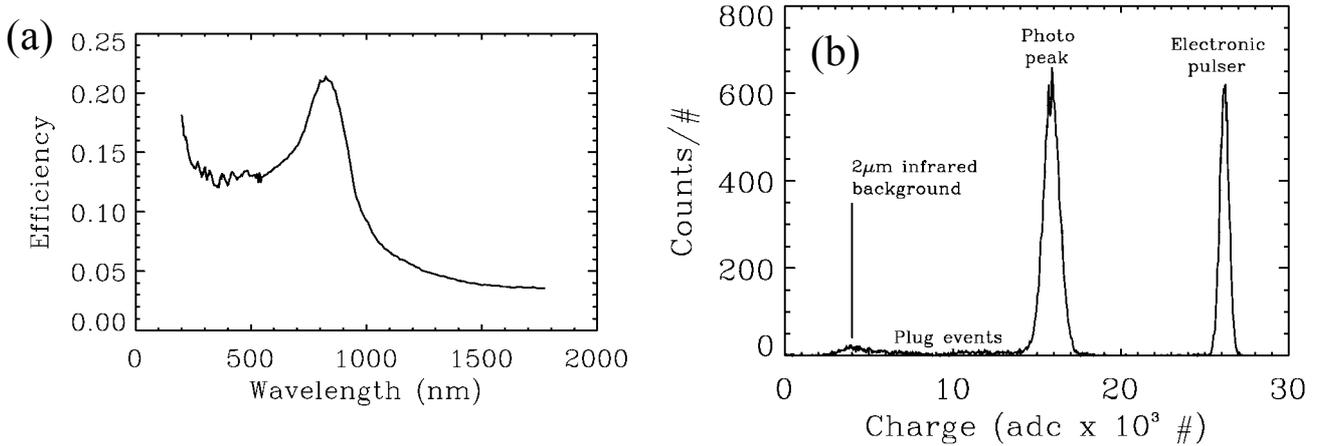

**Figure 7. (a)** Photon absorption efficiency for a 100 nm thick Al film on top of a 500 μm thick sapphire substrate as a function of the photon wavelength. The incidence angle of the incoming photons is normal to the surface. **(b)** Spectrum acquired with a 30 μm side length Nb-plugged Al junction under illumination with 500 nm photons (E = 2.48 eV). The applied bias voltage is 80 μV and the parallel magnetic field is equal to 32 Gauss.

An interesting feature in the IV-curves is the sudden current rise at a bias voltage of approximately 100 μV. As can be seen in Fig. 5(b), the sub-gap currents rise dramatically at bias voltages of 125, 110 and 100 μV for the 30, 50 and 70 μm side-length junctions respectively. In Figure 6, which shows the same IV-curves on a logarithmic scale, it can be seen that these current steps increase the sub-gap currents by almost three orders of magnitude. Well known mechanisms [10], such as Fiske resonances, multi-particle tunneling, self-coupling of Josephson radiation and multiple Andreev reflections, which are known to introduce a certain structure in the sub-gap currents of tunnel junctions, cannot explain this very drastic current step at a bias voltage that depends on the size of the junction. The current steps observed in these high quality, low $T_C$ and low loss junctions arise because of the very special non-equilibrium state, which forms due to the interplay of energy gain of the quasiparticles caused by sequential tunneling and energy loss due to down-scattering. During its lifetime a quasiparticle undergoes a large number of tunnel and back-tunnel processes. During every single tunnel or back-tunnel process the quasiparticle gains an energy $eV_b$. In addition, quasi-particle down-scattering to the gap energy is slow in low $T_C$ junctions because of the cubic dependence of the characteristic electron phonon scattering rate $\tau_0^{-1}$ on the $T_C$ of the material [11]. As a consequence, the quasi-particles in a biased low $T_C$ junction form a non-equilibrium energy distribution with quasiparticles at energies high above the gap energy. Note that it takes a certain amount of time for a quasiparticle to access the highest energy states, as it has to go through a cycle of subsequent tunnel, back-tunnel and down-scattering events. Therefore, in junctions with slow down-scattering, the highest energy state at which quasiparticles reside is limited by the loss time of the quasiparticles, which determines the maximum time available for a quasiparticle to go through the tunnel, back-tunnel and down-scattering cycle. Now, if the quasiparticles that reside at this highest energy level possess an energy that lies $2\Delta_g$ above the energy gap of the material, they will release a phonon of energy $2\Delta_g$, when relaxing down to the gap. This phonon can then break a Cooper pair and create two more quasiparticles. This process is called quasiparticle multiplication. These newly formed quasiparticles are then free themselves to undergo the tunnel, back-tunnel and down-scattering cycle. As a consequence, the number of quasiparticles in the electrode will be greatly enhanced and the tunnel current will rise sharply when the bias voltage is reached, which will lift the first quasiparticles into the active region. The fact that this threshold bias voltage is different for the three Al junctions of different size comes from the different loss times in the three junctions. As will be shown in the next section, the quasiparticle loss times are respectively equal to 20, 40 and 90 μsec in the 30, 50 and 70 μm junctions. Less time is available for the tunnel, back-tunnel and down-scattering cycle in the smaller junctions, which is the reason why the energy gained per tunnel event has to be larger in order for the quasiparticles to reach the active region within the loss time. Therefore, the threshold bias voltage at which the current step occurs is higher for the junctions with the faster losses.
Note that junctions without the Nb plug in the base electrode contact have much shorter quasiparticle loss times of the order of 5 μsec and do not show this very drastic current step in the IV-curve.
A very detailed discussion of this non-equilibrium phenomenon in high quality, low loss junctions is given in ref [12].

### 4.2. Optical photon detection

The different Nb-plugged junctions were illuminated through the sapphire substrate with monochromatic photons with a wavelength varying between 250 and 1000 nm. Figure 7(a) shows the absorption efficiency of a 100 nm thick Al film on a 500 μm thick sapphire substrate for incoming photon wavelengths ranging from 100 to 1800 nm [13]. For photons of wavelength varying between 300 and 1000 nm the efficiency roughly varies between 10 and 20%, the incomplete absorption efficiency in the thin Al film being completely due to reflections at the sapphire-Al interface. The intensity of the Xenon light source was adjusted in order to have about 100 counts per second in the base electrodes of the different detectors. Figure 7(b) shows a typical optical spectrum as acquired with the Nb-plugged Al junctions. The spectrum shown in the figure was taken with a 30 μm side length junction, which was illuminated with 2.48 eV photons ($\lambda$=500 nm). The bias voltage applied to the junction was 80 μV and the parallel magnetic field applied in order to suppress the Josephson current was equal to 32 Gauss. The peak in the middle of the figure represents the events absorbed in the Al junction. The peak on the right hand side of the figure is the electronic pulser peak, which measures the resolution broadening arising from the noise created by the charge sensitive pre-amplifier. The width of this peak can be quadratically subtracted from the photo peak width in order to yield the intrinsic resolution of the detector, cleared from any broadening effects by the electronics. Whereas the final measured resolution of the detector of course includes the broadening due to the electronics, the knowledge of the intrinsic resolution is still interesting, because it allows the comparison of the actual performance of the junction to theoretical predictions. On the left hand side of the spectrum a small number of events can be detected that arise from the ~2 μm infrared background created by the laboratory environment at 25°C. These thermal photons reach the detector through the optical fibre. In between the photo-peak and the microwave background peak an additional, very small number of events can be seen, which are events absorbed in the base lead and the plug of the detector.
The responsivity (electrons read out per eV of incoming photon energy) and decay time of the junctions were determined as a function of the different junction side lengths with an applied bias voltage of 100 μV. The result is shown in figure 8. Both the responsivity and the decay time of the detectors increase approximately proportional to the area of the detector, indicating that the major loss source is not in the bulk of the Al, but rather very localised, probably located at the contacts of the detector. One eV of photon energy creates approximately 3300 free charge carriers in the Al electrode, whereas the measured charge at the output of the detectors is much higher. The ratio of the charge measured at the output of the junction Q to the number of charge carriers initially created in the junction $Q_0$ is defined as the charge amplification factor $\bar{n} = Q/Q_0$ of the junction. This amplification factor $\bar{n}$ is rather large for our Al junctions. It is equal to 7 for the smallest junction (10 μm) and increases to 105 for the largest available junction size (70 μm). These large amplification factors increase the signal to noise ratio and are very helpful in reducing the electronic noise and increasing the resolving power of the detector. On the other hand the signal pulses are rather long. The decay time of the signal varies between 7 μsec for the smallest junction and 80 μsec for the largest junction. Of course, these long decay times limit the maximum count rate of the detector severely.
Note that junctions without the Nb plug in the base electrode contact have much shorter quasiparticle loss times of the order of 5 μsec and show therefore much reduced responsivity, which translates into a much lower S/N ratio.

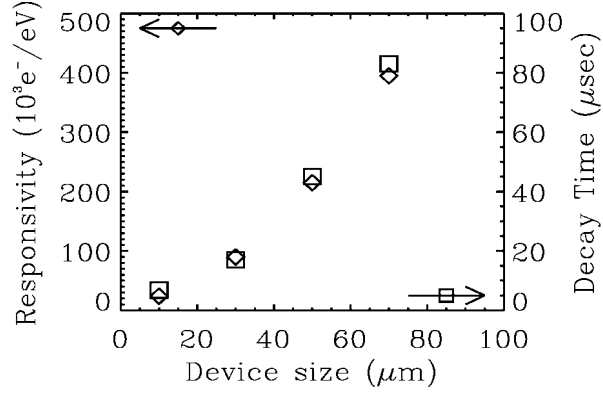

**Figure 8.** Responsivity and pulse decay time as a function of the device size of the different Nb-plugged Al junctions. The applied bias voltage is equal to 100 μV and the parallel magnetic field of the order of 30 Gauss.

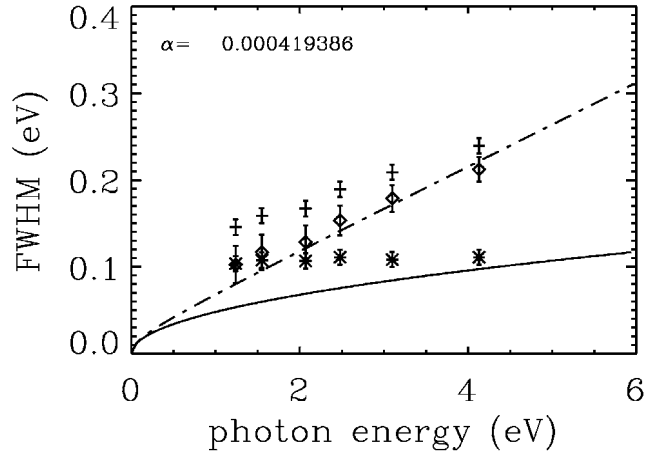

**Figure 9.** Measured energy resolution (crosses), measured electronic noise contribution (stars) and the derived intrinsic resolution (diamonds), as a function of incoming photon energy for the Nb-plugged Al device with 30 μm side length. The bias voltage applied is 95 μV. The solid line is the expected ideal energy resolution achievable. The dashed dotted line is a fit to the intrinsic resolution with the spatial broadening factor α used as fitting parameter.

Finally the resolution capabilities of the Nb-plugged detectors was determined as a function of the incoming photon energy. Figure 9 shows the measured photo-peak width, the electronic pulser width, which provides a monitor of the system noise, and the deduced intrinsic peak width as a function of incoming photon energy for the 30 μm side length device. The solid curve is the expected tunnel-limited resolution of Al STJs given by (1). The intrinsic peak width we observe is approximately a factor two broader than the expected tunnel limited resolution. The dashed-dotted line is a least-squares fit to the intrinsic resolution using a function of the form:

$$\Delta E = 2.355 \sqrt{1.7 \Delta_g (F+G)E + \alpha E^2} \,, \tag{2}$$

where α is a free fitting parameter representing inhomogeneities in the response across the junction. The deduced value for α is ~ 4 $10^{-4}$. This α value was deduced for all the available device sizes and is in all cases approximately equal to 4 $10^{-4}$. We recall here that the best Ta based junctions show values for α equal to 2 $10^{-6}$ [14] revealing that there is still considerable room for improvement.

## 5. TA-CONTACTED AL JUNCTIONS

As Ta is known for forming less metallic oxides than Nb [15], which behave as loss sources for the quasiparticles, we considered replacing the Nb used for fabricating the plugs and top contacts by Ta. This effectively should reduce losses at the contacts and homogenise the response of the detector considerably. Whereas this first assumption turned out to be correct, we encountered different problems related to the very long loss times of the quasiparticles in these Ta-plugged junctions.

### 5.1. I-V characterisation

Figure 10 shows the IV-curves of two 40 μm Al junctions with Ta contacts at 50 mK. One of the junctions possesses a Ta plug in the base electrode contact (dashed line), whereas the other does not posses a plug in the base electrode contact (solid line). Whereas the junction without the plug shows perfect leakage currents as low as 250 pA in the bias domain, the junction with the Ta plug in the base electrode contact shows more than three orders of magnitude higher currents. Both junctions are physically very close on the same chip and were operated during the same cooldown. This phenomenom is not a one off result, but is observed systematically for over 100 junctions on 12 different chips. In all cases the plugged junctions show subgap currents at 50 mK in excess of 100 nA, whereas the unplugged junctions show a perfect leakage current behavior with subgap currents well below a nA.

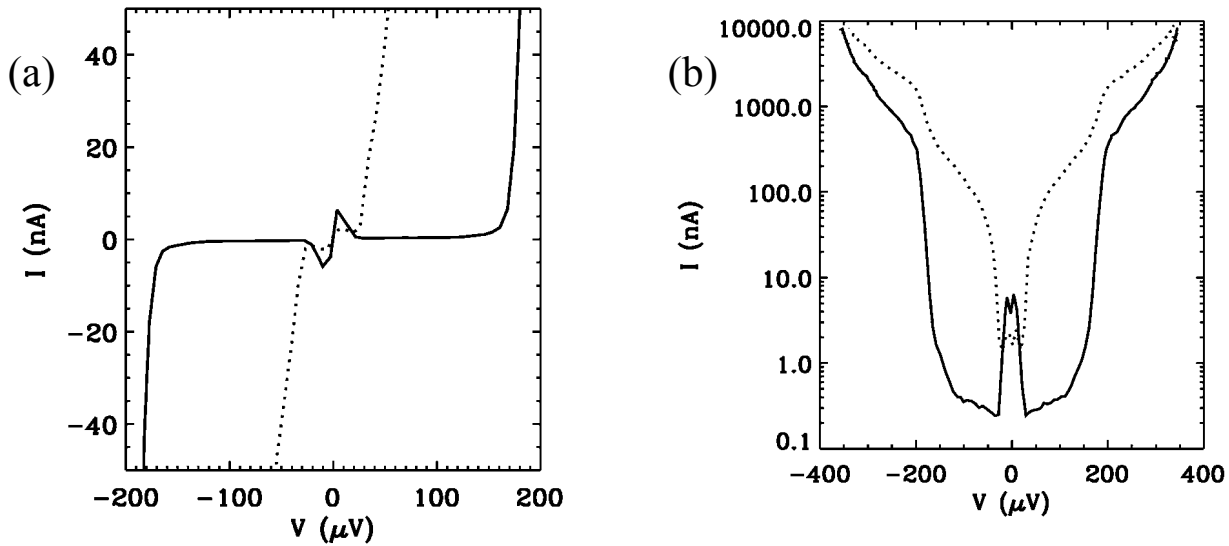

**Figure 10.** IV-curves of two 40 μm Al junctions with Ta contacts. **(a)** IV-curves on a linear scale in the bias domain varying from -200 to 200 μV. The dashed line shows the IV-curve of a junction with a Ta plug in the base electrode contact, whereas the solid line shows the IV-curve of a similar junction without the Ta plug in the base film. **(b)** Same IV-curves on a logarithmic scale showing the huge difference (more than 3 orders of magnitude) in subgap currents between the two junctions.

The reason for this difference in subgap current behavior is the much different quasiparticle loss time in both type of junctions. Whereas in the unplugged junctions the quasiparticles loss time is limited to approximately 10-15 μsec because of out-diffusion of the quasiparticles through the base electrode contact lead, the loss times in the plugged junctions are much longer, in excess of 150-200 μsec (see next section). When the quasiparticles live that long in the electrode, they have enough time to gain a lot of energy through multiple tunnel events, even at very low bias voltages, and start the quasiparticle multiplication procedure, which quickly increases the number of quasiparticles in the electrode to much larger numbers. Only when enough quasiparticles are present in the electrode in order to start quasiparticle losses by self-recombination, the growth of the quasiparticle population is stopped. Models describing the kinetics of the quasiparticle population already exist [12,16] and numerical simulations of the present data will be described in a future publication.

### 5.2. Optical photon detection

The different Ta-contacted junctions were also illuminated through the sapphire substrate with monochromatic photons. The unplugged Ta-contacted junctions show the same spatial broadening contribution to the photo-peak resolution as the Nb contacted junctions, because of the very localised losses to the base electrode contact lead. The Ta plugged junctions on the other hand are very difficult to bias, because of the very large subgap currents created by the non-equilibrium quasiparticle population. Only in the region 0-20 µV it is possible to bias the junction and to look at the response to optical photons. Whereas the spectra are extremely noisy, because of the high subgap currents, we could observe photo-pulses on the oscilloscope, which showed pulse decay times in excess of 150-200 µsec. This shows that the losses in the Ta-plugged Al junctions are extremely small and that quasiparticles created in the Al electrodes of the junctions life at least 150 to 200 µsec before they are lost.

In order to be able to operate these low-Tc, low-loss Al junctions, it will be necessary to introduce loss centres homogeneously over the area of the detector. This will on one hand reduce the non-equilibrium subgap currents created by quasiparticle multiplication and will on the other hand homogenise the response of the detector over the surface area and improve the energy resolution. Work is presently ongoing in order to realise such Al-based junctions with Ta leads and homogeneously distributed loss centres over the area of the detector.

## 6. CONCLUSIONS

We have successfully fabricated and operated very high quality Al STJs in a 50 mK environment. The junctions were illuminated with monochromatic optical photons with wavelengths varying between 250 and 1000nm. The junctions with Nb plugs in the base contact and Nb top contacts show charge amplification factors varying between 7 and 100, depending on the device size. The quasiparticle loss times of these junctions vary between 5 and 80 µsec and increase proportional to the area of the detectors, showing that all the losses are very localised and do not take place in the bulk of the Al, but rather in the Nb of the plugs and the top contacts. Because of this localised nature of the loss centres, the detectors show an energy resolution broadening due to an inhomogeneous gain over the area of the detector. The measured energy resolution is approximately a factor 2 larger than the best expected energy resolution for Al STJs.

As a consequence, Ta plugged and contacted Al junctions were fabricated, because Ta is known for its better properties with respect to quasiparticle loss centres, which should homogenise the detector gain with respect to the exact absorption position of the photon. Nevertheless, the Ta-contacted junctions are too perfect and the quasiparticles in these junctions have loss times in excess of 200 µsec. Because of these long loss times, even at low bias voltages, the quasiparticles gain a lot of energy via multiple tunnelling and start a multiplication procedure, which increases the initial quasiparticle population exponentially to much larger densities. As a consequence the IV-characteristics of the Ta-plugged junctions show subgap currents, which are three orders of magnitude larger than similar junctions without the Ta plug in the base electrode contact. These unplugged junctions do not prevent the out-diffusion of the quasiparticles out of the junction area through the base electrode contact and have therefore much shorter loss times of the order of 15 µsec. In order to solve this quasiparticle multiplication problem, quasiparticle loss centres have to be introduced homogeneously over the area of the detector. Work is presently ongoing to fabricate such junctions.

## 7. REFERENCES

[1] P. Verhoeve, "Low Temperature Detectors", 559, edited by F. S. Porter, D. McCammon, M. Galeazzi and C.K. Stahle, AIP Conf. Proc. No. 605, Melville, NY, (2002).
[2] A. G. Kozorezov, A. F. Volkov, J. K. Wigmore, A. Peacock, A. Poelaert, and R. den Hartog, Phys. Rev. B **61**, 11807 (2000).
[3] K. E. Gray, Appl. Phys. Lett. **32**, 392 (1978).
[4] M. Kurakado, Nucl. Instrum. Methods Phys. Res. A **196**, 275 (1982).
[5] N. Rando, A. Peacock, A. van Dordrecht, C. Foden, R. Engelhardt, B. G. Taylor, P. Gare, J. Lumley, C. Pereira, Nucl. Instrum. Methods Phys. Res. A **313**, 173 (1992).
[6] C. A. Mears, S. E. Labov, A. T. Barfknecht, Appl. Phys. Lett. **63**, 2961 (1993).
[7] D. J. Goldie, P. L. Brink, C. Patel, N. E. Booth, G. L. Salmon, Appl. Phys. Lett. **64**, 3169 (1994).
[8] P. Verhoeve, N. Rando, A. Peacock, D. Martin, R. den Hartog, Opt. Eng. **41** (6), 1170 (2002).
[9] http://www.vericold.de/adr.htm
[10 E. G. Wolf, *Principals of electron tunneling spectroscopy*, Oxford University Press, Oxford (1985).
[11] S. B. Kaplan, C. C. Chi, D. N. Langenberg, J. J. Chang, S. Jafarey, and D. J. Scalapino, Phys. Rev. B **14**, 4854 (1976).
[12] A. G. Kozorezov, J. K. Wigmore, A. Peacock, R. den Hartog, D. Martin, G. Brammertz, P. Verhoeve, N. Rando, Phys. Rev. B **69**, 184506 (2004).
[13] E. D. Palik, *Handbook of optical constants of solids*, Academic Press, New York (1985).
[14] G. Brammertz, P. Verhoeve, A. Peacock, D. Martin, N. Rando, R. den Hartog, D. J. Goldie, IEEE Trans. on Appl. Superc. **11** 1, 828, (2001).
[15] G. Brammertz, Ph.D. thesis , University of Twente , Enschede, The Netherlands, ISBN 90-365-1970-5 (2003).
[16] G. Brammertz, A.G. Kozorezov, J.K. Wigmore, R. den Hartog, P. Verhoeve, D. Martin, A. Peacock, A.A. Golubov, H. Rogalla , J. Appl. Phys. **94** , 5854 (2003).